\documentstyle[aps,epsfig]{revtex}
\begin{document}
\begin{titlepage}
\pagestyle{empty}
\baselineskip=18pt
\rightline{NBI--96--25}
\rightline{September, 1996}
\baselineskip=15pt
\vskip .2in
\begin{center}
{\large{\bf Many--Particle Correlations
in Relativistic Nuclear Collisions}}
\end{center}
\vskip .2truecm
\begin{center}
Henning Heiselberg 

{\it NORDITA, Blegdamsvej 17, DK-2100 Copenhagen \O., Denmark}

and Axel P.~Vischer

{\it Niels Bohr Institute,Blegdamsvej 17, DK-2100, Copenhagen \O, Denmark.}

\end{center}

\vskip 0.1in
\centerline{ {\bf Abstract} }
\baselineskip=15pt
\vskip 0.5truecm

Many--particle correlations due to Bose-Einstein interference are
studied in ultrarelativistic heavy--ion collisions.  We calculate the
higher order correlation functions from the 2--particle correlation
function by assuming that the source is emitting particles
incoherently. In particular parametrizations of and relations between
longitudinal, sidewards, outwards and invariant radii and
corresponding momenta are discussed. The results are especially useful
in low statistics
measurements of higher order correlation functions. 
We evaluate the three--pion correlation function recently measured by NA44
and predict the 2--pion--2--kaon correlation function. Finally,
many particle Coulomb corrections are discussed.
 
\end{titlepage} 

\baselineskip=15pt
\textheight8.9in\topmargin-0.0in\oddsidemargin-.0in

\section{Introduction}

Particle interferometry is an indispensable tool for the study of the
space-time structure of the emitting source created in
ultrarelativistic heavy--ion collisions. Recent experiments
~\cite{rama,sanjeev,NA35} not only enable us to obtain a good spatial
resolution in all 3 dimensions for the 2--pion correlation function,
but also provide large enough data samples to investigate many--pion
correlations~\cite{janus}.

The basic issue one wants to address and resolve with three or more
particle interferometry is, if there exist additional correlations
beyond Bose-Einstein (BE) interference of identical particles from an
incoherent source. In this paper we predict the three-- and more
particle correlation functions on these assumptions and explore
whether the two-body correlation function suffices to describe the
many--particle system.  In particular we compare to recent data on the
three-body correlation function. Any deviation found experimentally
from this prediction signals new physics like, e.g., coherent emission or
true higher order correlations. We use the measured two--particle
correlation functions as input and therefore obtain a parameter free
prediction of higher order correlation functions. 

Our starting point are the three--parameter fits to the two--pion
correlation function $C_{\rm 2}$. This correlation function depends on
the momentum difference, ${\bf q}$, of the pion pair.  In section 2 we
provide a simple method for calculating many-body correlation
functions due to Bose-Einstein interference between identical
particles emitted incoherently from a source and discuss to which
degree they can be determined from the two-body correlation function
alone.  To demonstrate our method, we reduce in section 3 the
dependence on the three--relative momenta to a dependence on only the
Lorentz invariant relative momentum and compare our result to the
corresponding one--parameter experimental fit and data~\cite{rama}.
The resulting expression is approximated by an expansion in the
deviations of the pion source from a spherical shape. We show that the
lowest order corrections are sufficient to reproduce the full
expression.

In section 4 we proceed to evaluate the three--pion correlation
function $C_{\rm3}$ and compare it with recent
results~\cite{janus}. We find that within the experimental errors the
prediction based only on three--pion correlations can reproduce the
data.  Finally in section 5 we discuss Coulomb effects and present a
simple way to correct many particle correlation functions.  We
conclude with a summary of our results.

\section{BE interference and many-body correlation functions}

We want to outline a general calculation of many-body
correlation functions which can be done elegantly for relativistic
heavy--ion collisions. The method reproduces in a simple calculation
the two-, three- and four-body correlation functions found previously by
diagrammatic techniques \cite{cramer}. Readers familiar with the results
may skip this section.

 The approximations, that we will employ, are:
\begin{itemize}
 \item {\it incoherent source}, i.e., particles are emitted independent of
each other and carry no phase information as is the case 
in (coherent) stimulated
emission. This is a strong assumption, which
can only be tested by comparison to experimental data. 
Existing two-body correlation functions
are well described by fully incoherent pion and kaon emission (see, e.g.,
\cite{henning}).
 \item {\it plane wave propagation} after emission. This assumes that
Coulomb repulsion can be corrected for (see, e.g., \cite{BB} and section 5)
and that other final state interactions are negligible. 
 \item {\it particle momenta are much larger than relative momenta}.
This is a good approximation as the
momenta of particles emerging from relativistic
heavy ion collisions are larger than the typical transverse momenta which
are $\sim 300$MeV/c. In comparison
the interference occur for relative momenta, $q$, less than the
inverse of the source size, $R\simeq 3-5$fm, and so 
$q \raisebox{-.5ex}{$\stackrel{<}{\sim}$} \hbar/R \sim 50$MeV/c.
 \item {\it large multiplicity}. The multiplicity in high energy nuclear
collision is hundreds or thousands of particles which is much larger than the
number of particles we want study correlations for.
\end{itemize}

The N-body correlation function relates directly to measured cross-sections
\begin{eqnarray}
    C(k_1,k_2,..,k_N) = \sigma^{N-1}
               \frac{d^{3N}\sigma/d^3k_1d^3k_2...d^3k_N}
                    {d^3\sigma/d^3k_1....d^3\sigma/d^3k_N}
\end{eqnarray}
where $k_1,...,k_N$ are the four-momenta of the N particles.

The wave function for a particle emitted at the source point $x_i$
with momentum $k_j$ is given by $\langle x_i|k_j\rangle=e^{ix_ik_j}$. 
We used here the assumption that the particle propagates like a plane wave.
We can then write down the symmetrized wavefunction for $N$ bosons,
\begin{eqnarray}
    |\psi_s(k_1,..,k_N)\rangle = 
    \frac{1}{\sqrt{N!}}\; 
     \sum_P  |P(k_1,...,k_N)\rangle  \;, \label{psi}
\end{eqnarray}
where $P$ is an operator representing all possible permutations of momenta
insuring that the wave function is symmetric. Anti-symmetrization in
case of fermions differs only by sign changes for odd permutations in
equation (\ref{psi}). The symmetrized wave
function for $2$ bosons is, for example, 
\begin{eqnarray}
|\psi_s(k_1,k_2)\rangle = 
    \frac{1}{\sqrt{2}}\; (\;|k_1,k_2\rangle  
	+|k_2,k_1\rangle)\;.
\end{eqnarray}
The symmetrized two-body wave-function in coordinate space is then
\begin{eqnarray}
\langle x_1,x_2|\psi_s(k_1,k_2)\rangle = 
    \frac{1}{\sqrt{2}}\; (e^{ik_1x_1}e^{ik_2x_2}+e^{ik_2x_1}e^{ik_1x_2})
    \;.
\end{eqnarray}

For an incoherent source, $S(x,k)$, emitting particles at space-time
point $x$ with momentum $k$, the correlation function can be written
as (see, e.g., \cite{book} for a derivation of the two-body case
also with a partially coherent source)
\begin{eqnarray}
    C_N(k_1,...k_N) = \int S(x_1,k_1)...S(x_N,k_N)
		|\langle x_1,...,x_N|\psi_s(k_1,...,k_N)\rangle|^2
                     d^4x_1.....d^4x_N \; , \label{CN}
\end{eqnarray}
when the source is normalized such that $\int S(x,k)d^4x=1$. 

Since $k_i\simeq K\equiv k_1+...+k_N)/N$ we can approximate
$S(x_i,k_i)\simeq S(x_i,K)$. We have furthermore permutation symmetry in 
the $x_i$'s in Eq. (\ref{CN}). As a consequence we obtain
for the wave function expectation value (for bosons) 
\begin{eqnarray}
|\langle x_1,...,x_N|\psi_s(k_1,...,k_N)\rangle|^2 &=& 
\frac{1}{N!} \sum_{P,P'}  \langle P'(k_1,...,k_N)|x_1,...,x_N\rangle
\;\langle x_1,...,x_N|P(k_1,...,k_N)\rangle  \nonumber\\
   &=& \sum_P e^{ix_1(k_1-P(k_{1}))}....e^{ix_N(k_N-P(k_{N}))} 
  \label{Permute}  \; .
\end{eqnarray}
We used here a symbolic notation. Every $P(k_{i})=k_j$ can be any of
the particle momenta. However, the set $P(k_{1}),...,P(k_{N})$ is the
same as the set $k_1,...,k_N$ - just permuted around in all possible
ways.  For the two-particle case
$N=2$ we obtain from Eq. (\ref{Permute}) $|\langle
x_1,x_2|\psi_s(k_1,k_2)\rangle|^2 =1+e^{i(x_1-x_2)(k_1-k_2)}$.  Note,
that the permutation symmetry in coordinates $x_1,...,x_N$ insures
that the permutations of either wave-function cancels the factor
$1/N!$.

From Eqs. (\ref{CN}) and (\ref{Permute}) we obtain
\begin{eqnarray}
   C_N(k_1,...k_N) = \sum_P \prod_{i=1}^N
    \int S(x,K) e^{ix(k_i-P(k_i))}
                     d^4x \; , \label{CNS}  
\end{eqnarray}
where the sum is over all the possible permutations and $P(k_{i})=k_j$
represents all possible momenta $k_j$ making up one permutation.
Here, we can for few-body systems ignore factors of $(n-N)/n$ due to
the large multiplicity $n$ in relativistic heavy ion collisions.

It is now convenient to define the Fourier transforms
\begin{equation}
    F(q)= \int dx\;S(x,K)\;\exp(iqx)\;,
                         \label{fs}
\end{equation}
where the dependence on the total momentum $K$ has been suppressed for
later convenience. The $N$-body correlation function is then
\begin{eqnarray}
   C_N(k_1,...k_N) = \sum_P \prod_{i=1}^N \; F(q_{ij}) \; , 
\label{CNF}  
\end{eqnarray}
where $q_{ij}=k_i-k_j=k_i-P(k_{i})$ and the sum is again over all possible
permutations.

It is now straight forward to write down the first few correlation functions.
The two-body
\begin{eqnarray}
   C_2(k_1,k_2) &=& 1+F(q_{12})F(q_{21})= 1+|F(q_{12})|^2 \; ; \label{cor2f}
\end{eqnarray}
and the three-body
\begin{eqnarray}
   C_3(k_1,k_2,k_3)
             =1&+&F(q_{12})F(q_{21})+F(q_{23})F(q_{32})+F(q_{31})F(q_{13})
                      \nonumber\\
                  &+&F(q_{12})F(q_{23})F(q_{31})+F(q_{21})F(q_{32})F(q_{13})
                      \nonumber\\
                =1&+&|F(q_{12})|^2+|F(q_{23})|^2+|F(q_{31})|^2
                     +2Re[F(q_{12})F(q_{23})F(q_{31})]   \;.  \label{cor3f}
\end{eqnarray}
For the four-body correlation function we give only the final result
\begin{eqnarray}
C_4(k_1,k_2,k_3,k_4) =1&+&|F(q_{12})|^2+|F(q_{13})|^2+|F(q_{14})|^2+
|F(q_{23})|^2+|F(q_{24})|^2+|F(q_{34})|^2 \nonumber \\
&+&2 \;(Re[F(q_{12})F(q_{23})F(q_{31})] + Re[F(q_{12})F(q_{24})F(q_{41})] 
\nonumber \\ &\;&\;\;+\;
Re[F(q_{13})F(q_{34})F(q_{41})] + Re[F(q_{23})F(q_{34})F(q_{42})] \;)
\nonumber \\
&+&2 \;(Re[F(q_{12})F(q_{23})F(q_{34})F(q_{41})] + 
Re[F(q_{12})F(q_{24})F(q_{43})F(q_{31})] 
\nonumber \\&\;&\;\;+\; 
Re[F(q_{13})F(q_{32})F(q_{24})F(q_{41})] \;) \nonumber \\
&+& |F(q_{12})|^2 |F(q_{34})|^2 + |F(q_{13})|^2 |F(q_{24})|^2 +
|F(q_{14})|^2 |F(q_{23})|^2\; .
\label{cor4f}
\end{eqnarray}
These expressions were also derived by a diagrammatic approach \cite{cramer}.

In case of two-pion and two-kaon correlations the
symmetrization/permutation should only be within each pair of
identical particles and so the 2x2 correlation function becomes
\begin{eqnarray}
C_{\rm 2 \times 2}({\bf q}_{\rm \pi_1 \pi_2},{\bf q}_{\rm K_1K_2}) =
\left(1+\;|F_\pi({\bf q}_{\rm \pi_1 \pi_2})|^2\right)
\left(1+\;|F_K({\bf q}_{\rm K_1K_2})|^2\right) \;.
\label{cor22f}
\end{eqnarray}
Since the pion and kaon sources may be different, $F_\pi$ and $F_K$ may
also differ.

The crucial assumption of an incoherent real source thus allows us to
write down any N-body correlation function if we know the function
$F(q)$.  Its norm can be extracted from the two-body correlation
function but not its phase.  If the Fourier transforms of the source
distributions Eq. (\ref{fs}) are complex, $F({\bf q}_{\rm ij})=|F({\bf
q}_{\rm ij})|\; e^{i\phi_{ij}}$, the two-body terms $|F({\bf q}_{\rm
ij})|^2$ in Eqs. (\ref{cor2f}) and (\ref{cor3f}) are insensitive to
the phase $\phi_{ij}$, whereas the three-body term $F({\bf
q}_{\rm12})F({\bf q}_{\rm 23})F({\bf q}_{\rm 31})$ in (\ref{cor3f})
generally will depend on the phase. However, translations in space and
time by {\it x} leads to a phase change $iq_{ij}x$ of $F({\bf q}_{\rm
ij})$ but due to momentum conservation, $q_{12}+q_{23}+q_{31}=0$, the
phase of $F({\bf q}_{\rm12})F({\bf q}_{\rm 23})F({\bf q}_{\rm 31})$ is
unchanged.  In general all correlation functions are invariant with
respect to translations.  Furthermore, if the source is symmetric,
i.e., $\rho(x)=\rho(-x)$, we find that $F({\bf q}_{\rm ij})$ are real
from their definition Eq. (\ref{fs}).  If the source is not symmetric
as, e.g.., when resonances contribute \cite{henning}, then the $F({\bf
q}_{\rm ij})$'s are generally complex and the phase in the last term
of Eq. (\ref{cor3f}) may carry interesting information.
The phase can, however, not be studied by two-body correlations alone
as seen from Eq. (\ref{cor2f}).

\section{Two--particle BE correlation function}

In this section we analyse the two-body correlation function in
relativistic nuclear collision and extract the crucial function
$|F(q)|$ for later use in many-body correlation functions. 
We will relate various parametrizations in particular as function of
the invariant relative momentum as will be important for the three-body
data.

A commonly used parametrization of the two--particle correlation function is the
three--parameter gaussian form
\begin{equation}
    C_{\rm 2}(q_{\rm l},q_{\rm s},q_{\rm o}) = 1 + \lambda_{\rm 2} \exp{(-q_{\rm l}^2 R_{\rm l}^2-q_{\rm s}^2
    R_{\rm s}^2-q_{\rm o}^2 R_{\rm o}^2)} ,
    \label{3dimc2}
\end{equation}
where the components of the relative momentum ${\bf q}={\bf p}_1-{\bf
p}_2$ are the longitudinal $q_l$ (along the beam axis), sidewards
$q_s$ (perpendicular to both the beam axis and the pair momentum) and
outwards $q_o$ (see Figure 1).

\vskip 0.5truecm
\begin{figure}
\centerline{
\psfig{figure=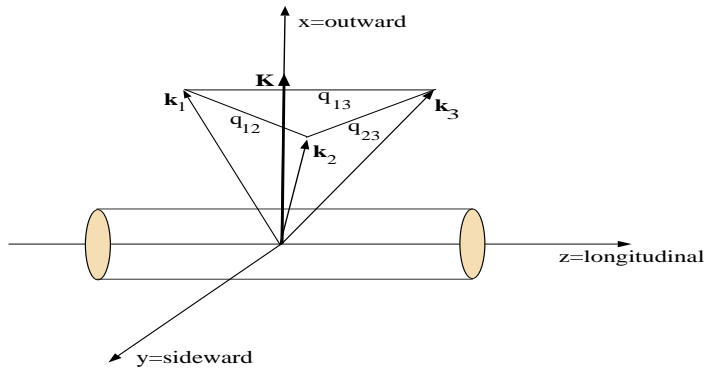,width=9.5cm,height=5cm}
}
\caption{
View of the interaction region for the case of 3 particle emission.}
\end{figure}

The dependence on $K$, or equivalently
the rapidity $y\simeq y_1\simeq y_2$ or transverse momentum
$K_\perp\simeq p_{\perp1}\simeq p_{\perp1}$, has been suppressed in
(\ref{3dimc2}). It is implicit in the source size parameters,
$R_{l,s,o}$.  Longitudinal expansion is revealed as a rapidity
dependent $R_l$ \cite{Sinyukov} while transverse expansion may be seen
through $R_{s,o}$ depending on the transverse momentum
\cite{Heinz,Csorgo}.  For an ideal chaotic source the parameter
$\lambda_{\rm 2}$ would be unity but a number of effects like a
partially coherent source, long lived resonances, final state
interactions and Coulomb screening effects can lead to a smaller value
for $\lambda_{\rm 2}$~\cite{henning}. Finally, we would like to point
out, that it is common to include a term $-2 q_{\rm o} q_{\rm l} R_{\rm ol}^2$ for
the parametrization in equation (\ref{3dimc2}). We did not include this term, since
we concentrate in this paper on the experimental data of NA44. The
generalisation of our calculations to include this term is straight forward.

Due to low statistics 
it is commonplace to perform a gaussian fit of one variable only
\begin{equation}
    C_{\rm 2}(q_{\rm inv}) = 1 + \lambda_{\rm inv} \exp{(-q_{\rm inv}^2
    R_{\rm inv}^2)},
    \label{1dimc2}
\end{equation}
where the invariant momentum squared is given by
\begin{eqnarray}
    q_{\rm inv}^2 &=& {\bf q}^2 - q_{\rm t}^2 
              = q_{\rm l}^2 + q_{\rm s}^2 +q_{\rm o}^2-q_{\rm t}^2 .
       \label{qinv}
\end{eqnarray}
Here, $q_{\rm t}=E_1-E_2\simeq {\bf K}\cdot{\bf q}/E_K$ since $q \ll K$.
Other variables, that also are used in parametrizations, are
$q^2_{R=\tau}=q^2_s+q^2_o+q^2_l+q^2_t$ or ${\bf
q}^2=q^2_s+q^2_o+q^2_l$. In the following we will perform an analysis
in the variable $q_{inv}$ but results can be taken over by changing
the sign of (for $q^2_{R=\tau}$) or removing (for ${\bf q}^2$) the
last term in equations (\ref{qinv}) and (\ref{qinva}).

Experimental data are often analysed in the
longitudinal center of mass system (LCMS) of each pair, in which $y=0$ and
${\bf K}$ is along the outward direction. In this system 
$q_{\rm t}=\beta_\perp q_o$, where 
\begin{equation}
    \beta_\perp = \frac{K_\perp}{\sqrt{m^2 + K_\perp^2}} ,
    \label{trans}
\end{equation}
is the transverse velocity in the LCMS. The LCMS--frame and the
different momenta are depicted in Figure 1. 
The invariant relative momentum reduces to
\begin{eqnarray}
    q_{\rm inv}^2 &\approx& q_{\rm l}^2 + q_{\rm s}^2 +q_{\rm o}^2\; (1-\beta_\perp^2).
    \label{qinva}
\end{eqnarray}

We can relate $C_{\rm 2}(q_{\rm inv})$ and $C_{\rm 2}(q_{\rm l},q_{\rm s},q_{\rm o})$
if both are determined from the same data sample\\
\begin{eqnarray}
C_{\rm 2}(q_{\rm inv}) \simeq \frac{\int d^3 q\; C_{\rm 2}(q_{\rm l},q_{\rm s},q_{\rm o})\;
\delta(q_{\rm inv}^2 - q_{\rm l}^2 - q_{\rm s}^2 -q_{\rm o}^2\; (1-\beta_\perp ^2))}
{\int d^3 q\; \delta(q_{\rm inv}^2-q_{\rm l}^2-q_{\rm s}^2-q_{\rm o}^2 \;(1-\beta_\perp^2))}.
\label{rel2}
\end{eqnarray}
The $\delta$--function enforces equation (\ref{qinv}),
while the numerator assures proper normalization. 

In high energy nuclear collisions the sources sizes $R_i$, {\it i=l,s,o}, 
are often similar in size as 
in high energy pp, pA and S+Pb collisions 
(see \cite{rama,sanjeev} or Tables 1 and 2).
In such cases we can expand in the difference between the radii and
obtain from (\ref{rel2})
\begin{eqnarray}
C_{\rm 2}(q_{\rm inv}) = 1 + \lambda_{\rm 2} \exp{(-q_{\rm inv}^2 \bar{R}^2)}\;
\left[1+\frac{1}{15}\;q_{\rm inv}^4 \;\sum_{i=l,s,o'} (R_i^2-\bar{R}^2)^2
  + {\cal O}(q_{\rm inv}^6(R_i^2-\bar{R}^2)^3) \right].
\label{c2appro}
\end{eqnarray}
where
\begin{equation}
\bar{R}^2 = \frac{1}{3} (R_{\rm l}^2+R_{\rm s}^2 +{R_{\rm o'}}^2),
\label{ri}
\end{equation}
is the average source size squared and 
\begin{equation}
   R_{\rm o'}=R_{\rm o}/\sqrt{1-\beta_\perp^2} 
\end{equation}
is the outward radius boosted by the transverse velocity. 
In the limit of small $(R_i^2-R_{\rm inv}^2)$ or small $q_{\rm inv}$
we recover $R_{\rm inv}=\bar{R}$ from a
comparison of (\ref{c2appro}) and  (\ref{1dimc2}).
Notice that $\bar{R}$ is defined such that 
the leading correction in (\ref{c2appro}) is second order
in $(R_i^2-R_{\rm inv}^2)$ and fourth order in $q_{\rm inv}$.  

Defining the transverse and longitudinal source deformation parameters 
\begin{eqnarray}
\delta_\perp &=& \;\frac{R_{\rm o'}^2 -R_{\rm s}^2}{2\bar{R}^2} \nonumber \\
\delta_{\rm l} &=& \;\frac{R_{\rm l}^2 - \bar{R}^2}{2\bar{R}^2}
\label{ecc}
\end{eqnarray}
we can rewrite equation (\ref{c2appro}) as
\begin{eqnarray}
C_{\rm 2}(q_{\rm inv}) = 1 + \lambda_{\rm 2} \exp{(-q_{\rm inv}^2\bar{R}^2)}\;
     \left[1+\frac{2}{5}\;q_{\rm inv}^4\bar{R}^4\;
     (\frac{1}{3}\;\delta_\perp^2+\delta_{\rm l}^2) 
  + {\cal O}(q_{\rm inv}^6\bar{R}^6\delta^3 )\right] .
\label{c2approx}
\end{eqnarray}
The definitions of these deformations or eccentricities are related to
the deformation 
parameters $\beta$ and $\gamma$ of Bohr \& Mottelson~\cite{BM}.

The deformations have a physical meaning. In many cylindrical symmetric
models for sources with an emission time $\delta\tau$, one finds that
$R_o^2-R_s^2=\beta_\perp^2\delta\tau^2$ \cite{Heinz,Csorgo,henning}.
Thus
\begin{equation}
 \delta\tau^2 =
\bar{R}^2 \;[\delta_\perp\;(2\beta_\perp^{-2}-1) +\delta_{\rm l}-1] .
\end{equation}
In the Bjorken 1-dimensional hydrodynamic expansion model, one finds
to a good approximation that $R_{\rm l}^2=T\tau^2/\sqrt{m^2+K_\perp^2}$
in LCMS, where $\tau$ is the average invariant time
it takes to freeze-out after the collisions and T is the temperature at
freeze-out. In that case
\begin{equation}
\tau^2 = \bar{R}^2\;\frac{p_\perp}{\beta_\perp\;T}\; (2\delta_{\rm l}\;+1)\;.
\end{equation}

It may be a coincidence that the source appears to be
spherical\footnote{ By spherical we mean $R_l=R_s=R_o'$ but not
necessarily a spherical and isotropic source}. Rephrased in terms of
the deformations, we can say that the emission time is small and that
the freeze-out happens when longitudinal expansion is about as large
as the transverse size of the system, {\it i.e.} when the source
expands in three dimensions.

We applied these results to the data samples taken in the experiment
NA44~\cite{rama,sanjeev}. Figure 2 shows in the upper left corner the 
results for the reaction
$S+Pb \rightarrow \pi^+ \pi^+ X$, while the upper right corner
corresponds to the
reaction $p+Pb \rightarrow \pi^+ \pi^+ X$.  In the two plots in the
bottom row of Figure 2 we
performed the analogous calculations for kaons. The lower left shows the
results for the reaction $S+Pb \rightarrow K^+ K^+ X$, while lower right
corresponds to the reaction $p+Pb \rightarrow K^+ K^+ X$. A
three-parameter fit was applied to the data samples with the results
listed in Table 1 and Table 2.

\vskip 0.5truecm

\begin{center}
\begin{tabular}
{|c|c|c|c|c|c|}\hline
\,&$\lambda_{\rm 2}$&$R_{\rm l}\; (fm)$&$R_{\rm s}\; (fm)$&$R_{\rm o}\; (fm)$&
$\langle p_\perp\rangle\;(MeV)$ \\ \hline
S+Pb&0.56$\pm$0.02&4.73$\pm$0.26&4.15$\pm$0.27&4.02$\pm$0.14 &
150 \\ \hline
p+Pb&0.41$\pm$0.02&2.34$\pm$0.36&2.00$\pm$0.25&1.92$\pm$0.13 &
150 \\ \hline
\end{tabular}
\end{center}
{\baselineskip=17pt
\noindent {\bf Table 1.} Parameters of the three--parameter fits to the
NA44 data sample at 200 AGeV 
for the two--pion correlation function (from~\cite{rama}).}
\vskip 0.5truecm
\baselineskip=20pt

\begin{center}
\begin{tabular}
{|c|c|c|c|c|c|}\hline
\,&$\lambda_{\rm 2}$&$R_{\rm l}\; (fm)$&$R_{\rm s}\; (fm)$&$R_{\rm o}\; (fm)$&
$\langle p_\perp\rangle\;(MeV)$ \\ \hline
S+Pb&0.82$\pm$0.04&3.02$\pm$0.20&2.55$\pm$0.20&2.77$\pm$0.12 &
246 \\ \hline
p+Pb&0.70$\pm$0.07&2.40$\pm$0.30&1.22$\pm$0.76&1.53$\pm$0.17 &
237 \\ \hline
\end{tabular}
\end{center}
{\baselineskip=17pt
\noindent {\bf Table 2.} Parameters of the three--parameter fits to the
NA44 data sample at 200 AGeV 
for the two--kaon correlation function (from~\cite{sanjeev}).}
\vskip 0.5truecm
\baselineskip=20pt

\begin{figure}
\centerline{
\mbox{
\psfig{figure=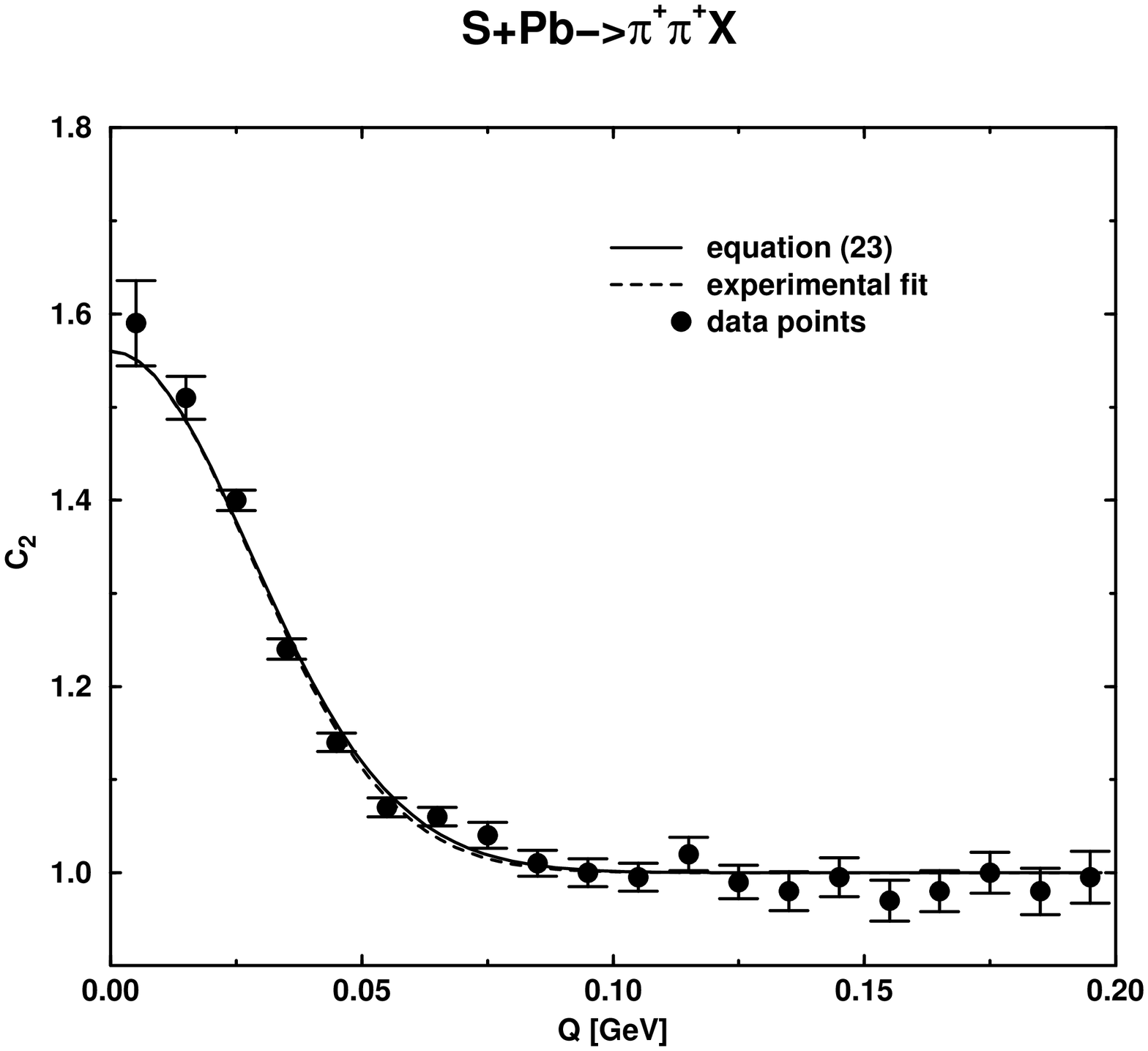,width=9.5cm,height=5cm,angle=-90}
\psfig{figure=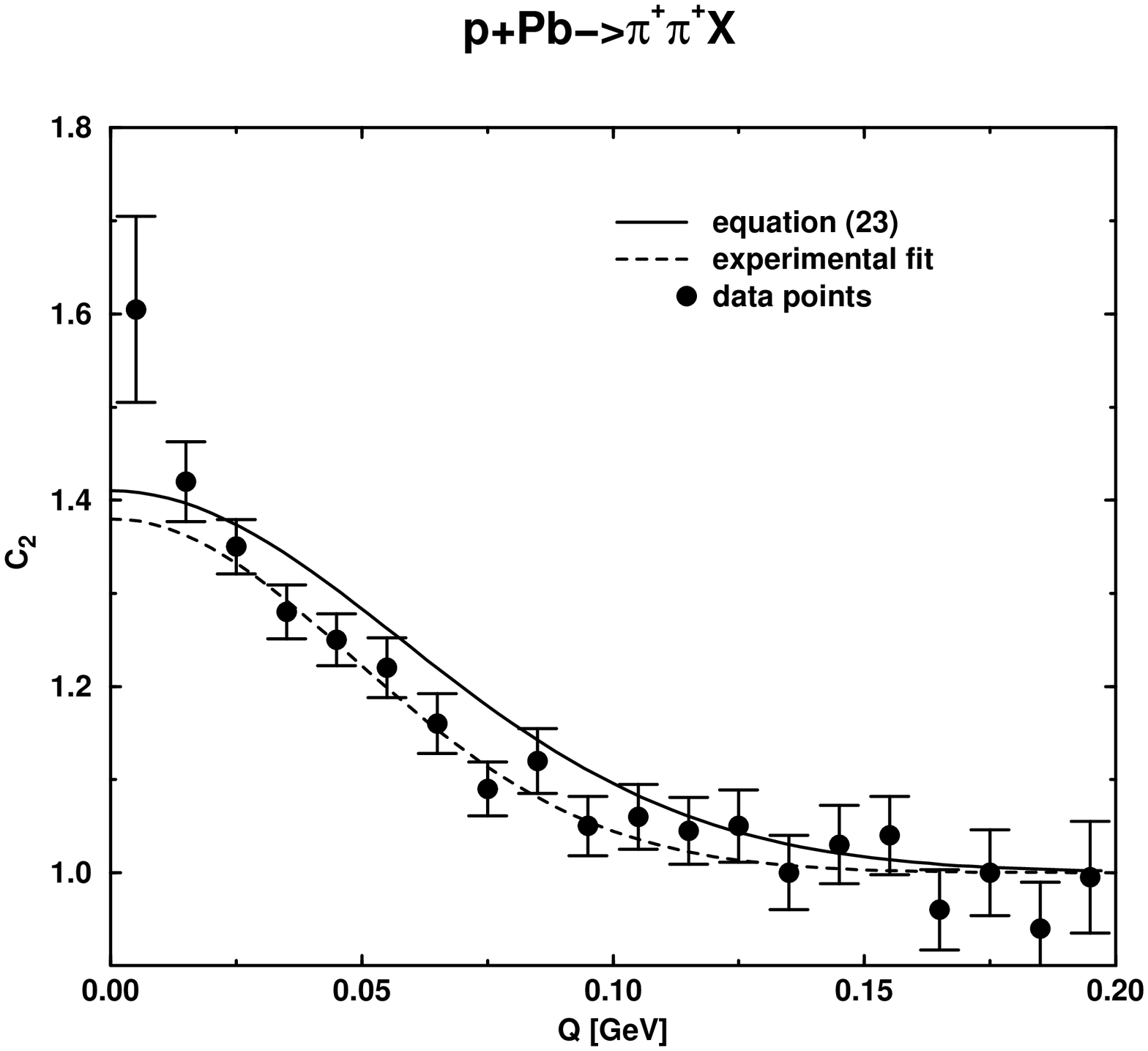,width=9.5cm,height=5cm,angle=-90}
}}
\centerline{
\mbox{
\psfig{figure=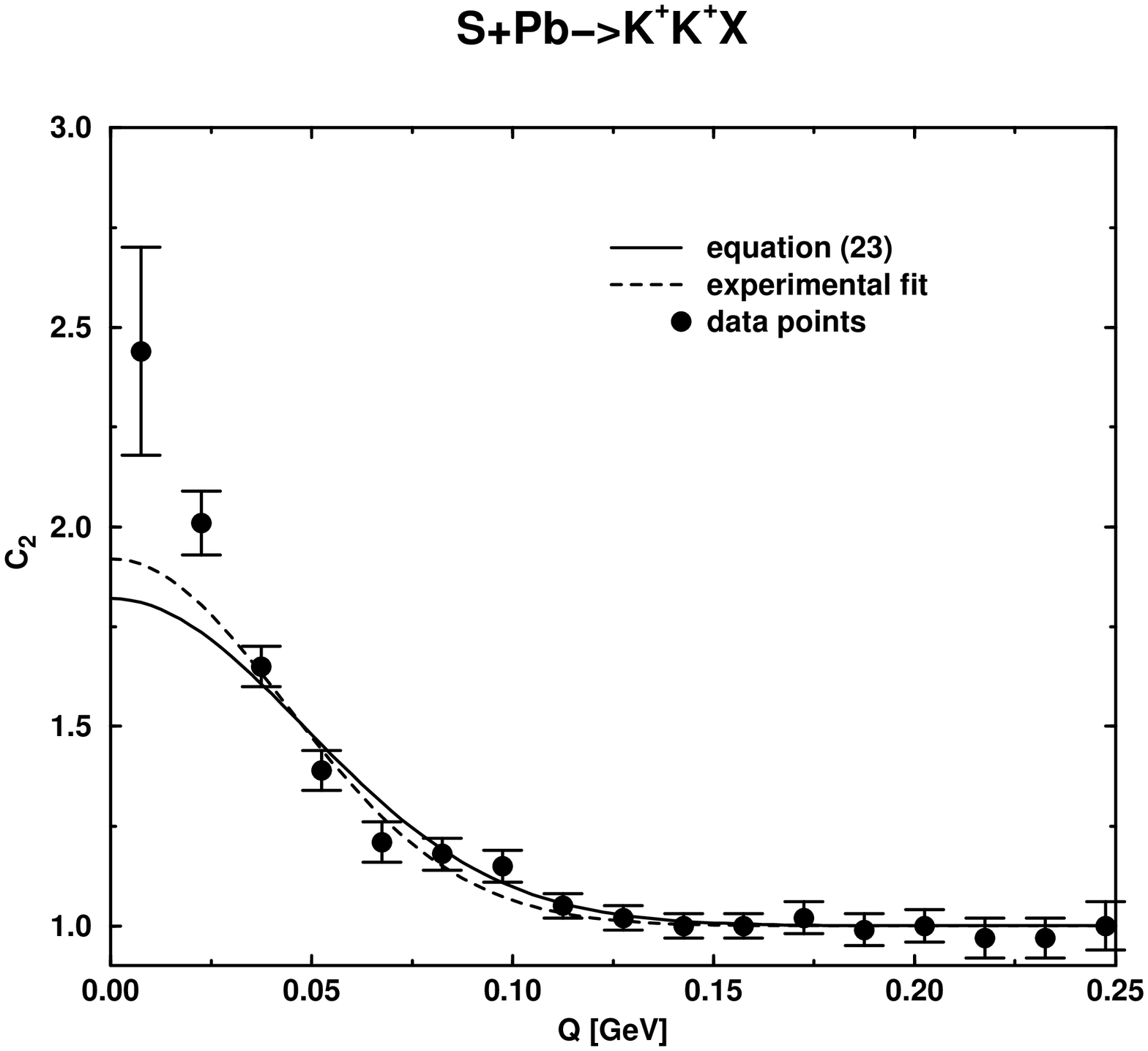,width=9.5cm,height=5cm,angle=-90}
\psfig{figure=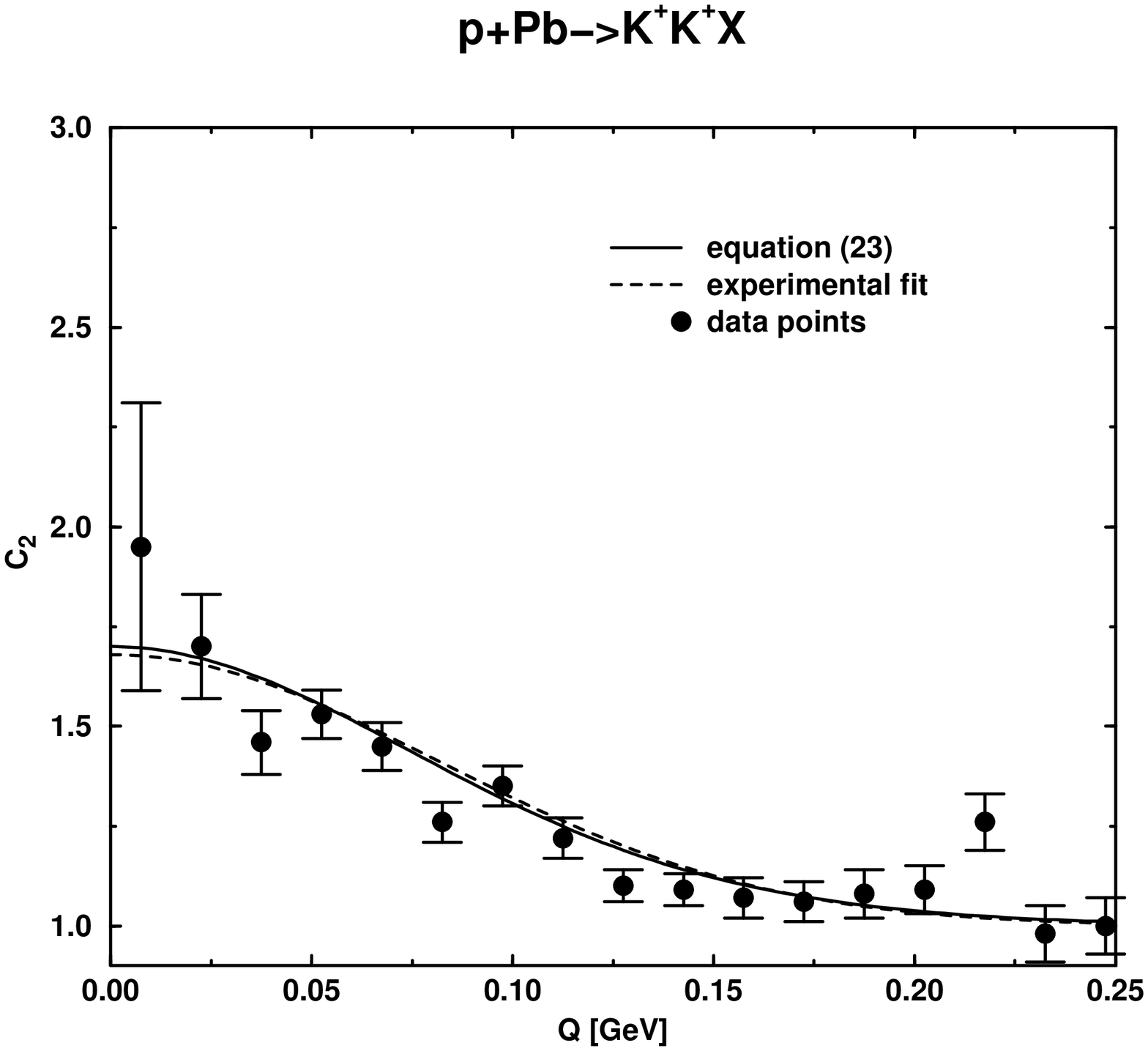,width=9.5cm,height=5cm,angle=-90}
}}
\caption{ Two--pion correlation function for the reaction $S+Pb \rightarrow
\pi^+ \pi^+ X$ (upper left), $p+Pb \rightarrow \pi^+ \pi^+ X$ (upper
right), $S+Pb \rightarrow K^+ K^+ X$ (lower left) and 
$p+Pb \rightarrow K^+ K^+ X$ (lower right). The 
data points are from NA44 \protect{\cite{rama,sanjeev}}, the 
solid line is calculated using (\protect{\ref{c2approx}}) and the dashed line 
is the 1--parameter fit (Table 5 and 6). }
\end{figure}

We would like to mention that to apply
equation (\ref{rel2}) and (\ref{c2approx}) to the three-parameter fits
we have for some data samples to take into account the experimental 
cut--offs for low
momenta. In some instances a bin size of $\delta q \sim 10$ MeV/c was
applied for each momentum projection and the corresponding 
two other momentum projections were summed
over all momenta between $0$ and $q_{\rm cut} \sim 20$ MeV/c. This
prescription modifies the value
of $\lambda_{\rm 2}$. For the longitudinal direction we find, for
example,
\begin{eqnarray}
C_{\rm 2}(q_{\rm l}) &=& \int_{\rm 0}^{q_{\rm cut}} \frac{dq_{\rm s}}{q_{\rm cut}} \;
\int_{\rm 0}^{q_{\rm cut}} \frac{dq_{\rm o}}{q_{\rm cut}}\;
C_{\rm 2}(q_{\rm l},q_{\rm s},q_{\rm o}) \nonumber \\
&=& 1 + \lambda_{\rm 2} \exp(-q_{\rm l}^2 R_{\rm l}^2)\;  \;\frac{\pi}{4} \;
\frac{{\rm Erf}(R_{\rm s} \;q_{\rm cut})}{R_{\rm s} \;q_{\rm cut}}\;
\frac{{\rm Erf}(R_{\rm o} \;q_{\rm cut})}{R_{\rm o} \;q_{\rm cut}} \nonumber \\
&\simeq& 1 + \lambda_{\rm 2} \exp(-q_{\rm l}^2 R_{\rm l}^2)\; 
        (1-\frac{1}{3}q_s^2R_s^2)( (1-\frac{1}{3}q_o^2R_o^2) ,
          \quad q_{\rm cut}R_i\ll 1 ,
\label{corr}
\end{eqnarray}
where $\rm Erf(x)$ 
is the error
function. Corresponding equations apply to the other two directions and
result in an apparent reduction of the experimentally measured
$\lambda_{\rm 2}$. 
In the case of NA44 this prescription was not applied in the fitting
procedure~\cite{priv}.
 
In Figure 2 we find reasonably good agreement between the
data points (black dots), the
1--parameter fit (dashed line) and our calculation using equation 
(\ref{c2approx}) (solid line). The approximation (\ref{c2approx}) is
in this case hardly distinguishable from the exact expression
(\ref{rel2}). The parameters used for the upper row of Figure 2
are listed in Table 3,
while the parameters used for the lower row
in Table 4.
The parameters for the 1--parameter fit are listed in 
Table 5 and 6 for pions and kaons respectively.   

\vskip 0.5truecm

\begin{center}
\begin{tabular}
{|c|c|c|c|c|c|}\hline
\,&$R_{\rm inv} \;(fm)$&$\langle\beta_\perp\rangle$&
$\delta_{\rm l}$&$\delta_\perp$ \\ \hline
S+Pb&4.98&0.73&-0.05&0.35 \\ \hline
p+Pb&2.41&0.73&-0.03&0.34 \\ \hline
\end{tabular}
\end{center}
{\baselineskip=17pt
\noindent {\bf Table 3.} Source parameters for the two--pion correlation
function determined according to equations (\ref{trans}) and (\ref{ecc}).}
\vskip 0.5truecm
\baselineskip=20pt

\begin{center}
\begin{tabular}
{|c|c|c|c|c|c|}\hline
\,&$R_{\rm inv} \;(fm)$&$\langle\beta_\perp\rangle$&
$\delta_{\rm l}$&$\delta_\perp$ \\ \hline
S+Pb&2.90&0.44&0.04&0.18 \\ \hline
p+Pb&1.84&0.43&0.35&0.20 \\ \hline
\end{tabular}
\end{center}
{\baselineskip=17pt
\noindent {\bf Table 4.} Source parameters for the two--kaon correlation
function determined according to equations (\ref{trans}) and (\ref{ecc}).}
\vskip 0.5truecm
\baselineskip=20pt

\begin{center}
\begin{tabular}
{|c|c|c|}\hline
\,&$\lambda_{\rm inv}$&$R_{\rm inv} \;(fm)$\\ \hline
S+Pb&$0.56\pm0.03$&$5.00\pm0.22$ \\ \hline
p+Pb&$0.38\pm0.03$&$2.89\pm0.30$ \\ \hline
\end{tabular}
\end{center}
{\baselineskip=17pt
\noindent {\bf Table 5.} Parameters of the 1--parameter fit to the
NA44 data sample for the two--pion correlation function (from~\cite{rama}).}
\vskip 0.5truecm
\baselineskip=20pt

\begin{center}
\begin{tabular}
{|c|c|c|}\hline
\,&$\lambda_{\rm inv}$&$R_{\rm inv} \;(fm)$\\ \hline
S+Pb&$0.92\pm0.08$&$3.22\pm0.20$ \\ \hline
p+Pb&$0.68\pm0.06$&$1.71\pm0.17$ \\ \hline
\end{tabular}
\end{center}
{\baselineskip=17pt
\noindent {\bf Table 6.} Parameters of the 1--parameter fit to the
NA44 data sample for the two--kaon correlation function (from~\cite{rama}).}
\vskip 0.5truecm
\baselineskip=20pt


\section{Three-particle BE correlation functions}

We can now apply the results of the previous sections to predict all three--
and more particle correlation functions. 
The three-pion correlation function recently determined by NA44~\cite{janus}  
is especially relevant. A 1--parameter Gaussian fit was applied to the
data and the radius and $\lambda$-parameter of this fit are listed in

\vskip 0.5truecm
\begin{center}
\begin{tabular}
{|c|c|c|}\hline
\,&$\lambda$&$R_{\rm inv} \;(fm)$\\ \hline
S+Pb&$2.1\pm0.4$&$2.7\pm0.2$ \\ \hline
\end{tabular}
\end{center}
{\baselineskip=17pt
\noindent {\bf Table 7.} Parameters of the 1--parameter fit to the
NA44 data sample for the three--pion correlation function (from~\cite{janus}).}
\vskip 0.5truecm
\baselineskip=20pt

Table 7. The data was fitted in terms of the sum, $Q_{\rm 3}$, of the
relative invariant momenta of the 3 particles labeled 1,2,3
\begin{equation}
   Q_{\rm 3}^2 = q_{\rm 12}^2 + q_{\rm 13}^2 + q_{\rm 23}^2\;.
   \label{q3}
\end{equation}
The relative invariant momentum of particle i and j is hereby given as
in equation (\ref{qinv}) (see also Figure 1)
\begin{equation}
   q_{\rm ij}^2 = {\bf q}_{\rm ij}^2 - q_{\rm t,\, ij}^2 \; .
\end{equation}
We proceed to evaluate the three--pion correlation function $C_3(Q_3)$, 
starting from 
\begin{eqnarray}
C_{\rm 3}(Q_{\rm 3}) = \frac{
\int d{\bf q}_{\rm 12}\;d{\bf q}_{\rm 13}\;d{\bf q}_{\rm 23}\; 
C_{\rm 3}({\bf q}_{\rm 12},{\bf q}_{\rm 13},{\bf q}_{\rm 23})
\;\delta(Q_{\rm 3}^2-q_{\rm 12}^2-q_{\rm 13}^2-q_{\rm 23}^2 )
\;\delta^3({\bf q}_{\rm 12}+{\bf q}_{\rm 13}+{\bf q}_{\rm 23})}
{\int d{\bf q}_{\rm 12}\;d{\bf q}_{\rm 13}\;d{\bf q}_{\rm 23}\; 
\delta(Q_{\rm 3}^2-q_{\rm 12}^2-q_{\rm 13}^2-q_{\rm 23}^2)\;
\delta^3({\bf q}_{\rm 12}+{\bf q}_{\rm 13}+{\bf q}_{\rm 23})}.
\label{rel3}
\end{eqnarray}
The $\delta$--function of the sum of the relative
momentum vectors assures that the 3 particles span a triangle.

As mentioned we cannot extract the phase from two-body correlation function.
We shall therefore in the following assume that $F({\bf q}_{\rm ij})$ 
are real functions which then are known from the two-body correlation function.
The three-body and any many-body correlation function is then completely
given by equation (\ref{cor3f}) and (\ref{CNF}).

At present statistics do not allow experimentalists to map out the 
three-body correlation function of all six variables (nine including
${\bf K}$) and we therefore reduce it according to Eq. (\ref{rel3}).

In the case of a spherical source, we obtain
from  (\ref{rel3}) and (\ref{cor3f}) 
\begin{eqnarray}
C_{\rm 3}(Q_{\rm 3}) = 1
  &+& 3\;\lambda_2\;\exp(-x)\;
  \tilde{I}_1(x) 
   + 2\;\lambda_2^{1.5}\exp(-y)\;,
 \label{c3approx}
\end{eqnarray}
and for a slightly deformed source, we obtain
\begin{eqnarray}
C_{\rm 3}(Q_{\rm 3}) = 1
&+& 3\;\lambda_2\;\exp(-x)\;
\left[\tilde{I_1}(x) + \frac{x^2}{15}\;
\;(\frac{1}{3}\;\delta_\perp^2+\delta_{\rm l}^2)\; 
\left( 4\;\tilde{I_1}(x)-3\;\tilde{I_2}(x)\right)
\right]\nonumber\\
&+& 2\;\lambda_2^{1.5}\exp(-y)\;
\left[ 1+\frac{y^2}{10}\;
(\frac{1}{3}\;\delta_\perp^2+\delta_{\rm l}^2)\right]
+ O(\bar{R}^6Q_{\rm 3}^6 \delta^3)\;.
\label{c3appros}
\end{eqnarray}
where we used the dimensionless variable $x=\bar{R}^2\;Q_{\rm 3}^2/3$ and
$y= \bar{R}^2\;Q_{\rm 3}^2/2$.
$\tilde{I}_{\nu}$ is related to the standard Bessel function $I_{\nu}$
\begin{eqnarray}
 \tilde{I_{\nu}}(x)&=&(\frac{2}{x})^{\nu}\;I_{\nu}(x) = 
\sum_{n=0}^\infty \frac{(x/2)^{2n}}{n!(n+\nu)!}\;. \nonumber
\end{eqnarray}

Figure 3 depicts this result, together with a 1-parameter Gaussian
fit of reference~\cite{janus}. A satisfactory agreement is
achieved. Based on the currently available data we seem not to need
any correlations beyond the two-pion correlations.

\begin{figure}
\centerline{
\psfig{figure=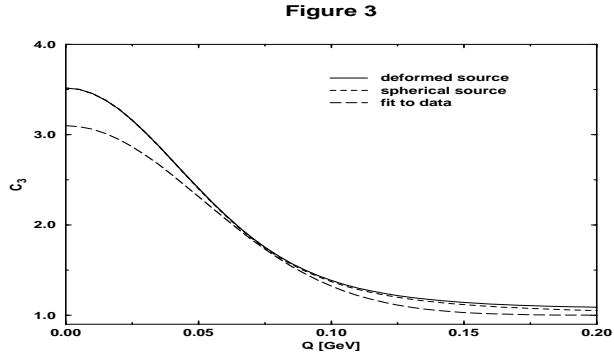,width=9.5cm,height=5cm,angle=-90}}
\caption{
Three--pion correlation function for the
reaction $S+Pb \rightarrow \pi^+ \pi^+ \pi^+  X$. The data points are
from NA44~\protect{\cite{janus}}, the solid line is 
calculated using (\protect{\ref{c3approx}}), assuming a spherical source 
(no deformations) and the
dashed--dotted line uses the
approximation (\protect{\ref{c3approx}}), allowing for deformations. 
Finally, the dashed line is the 1--parameter fit
(Table 7).  }
\end{figure}

The three-body correlation function of equation (\ref{c3approx}) or 
(\ref{c3appros}) is
a sum of two gaussians.
However, the current experimental
resolution cannot resolve the last term in these expressions,
since it decays with an 3/2
times larger exponential than the 3 explicit two-particle contributions and
the data does not explore sufficiently small $Q_3$. More statistics is
needed to resolve this second term which may be available in Pb+Pb collisions
at the SPS or future RHIC or LHC heavy ion collisions. It is the second
term that may carry a phase if, e.g., the source is asymmetric. It
would therefore be most interesting to study three and more particle
correlations at small $Q_3$ to learn more about the source.

The method can be generalized to study different aspects
of the 9 dimensional phase space, if the experimental resolution should
permit so, or extended to calculate any many--particle correlation
function of invariant momenta. 

As a last example, we would like to proceed to the 4-particle
correlation function. Current data samples are not large enough to
study pure 4--pion correlation functions but they do allow us to study
for example two--pion--two--kaon correlation functions $C_{2\times2}$
of equation (\ref{cor22f}). As described above the two--particle
correlation studies for pions and kaons can be used to obtain the
invariant radii and eccentricities for the two particles and
subsequently employing equations (\ref{c2approx}) and (\ref{cor2f}) allows
us to calculate $C_{\rm 2 \times 2}(q_{\rm \pi \pi},q_{\rm KK})$. If
the data should show any deviation from this result, we would have
evidence for cross-correlations between pions and kaons.

\section{Coulomb Corrections}

In a recent paper~\cite{BB} Baym and Braun--Munzinger show that the
standard use of the Gamow correction for Coulomb repulsion between
pions of like charge is incorrect. The relevant length scale for the
two pion wavefunction is their classical turning point in their
Coulomb field, $\sim 4\alpha m_\pi/q^2\raisebox{-.5ex}{$\stackrel{<}{\sim}$} 2$ fm, for relative momenta
$q\raisebox{-.5ex}{$\stackrel{>}{\sim}$} 10$ MeV/c, and not the two-pion Bohr radius,
$a_0=2/m_\pi\alpha=387$ fm. Since the turning point is smaller than the
typical source size in high energy nuclear collisions, the two-body
interactions are insignificant as compared to those of the source.  It
is instead a better approximation simply to assume classical
trajectories for the two charged particles in their Coulomb field,
$V_C(r)=Z_1Z_2e^2/r$, once they are separated by a distance $r_0$ of
order the source size.  A good description of $\pi\pi$ and $\pi p$
correlation data from nuclear collisions at AGS was obtained 
with an initial separation of $r_0\simeq 9$ fm. 

The simple model of Ref. \cite{BB} can be generalized to N particles.
Energy conservation implies that the final energies, $E_i$, of the
particles are related to the initial ones, $E_{0,i}$, by
\begin{eqnarray}
  \sum_{i=1}^N E_i = \sum_{i=1}^N E_{0,i} \,+\, \frac{N(N-1)}{2}
\frac{e^2}{R_0} \label{ec}\,,
\end{eqnarray}
where $R_0$ is the N-particle equivalent of $r_0$ and corresponds to
an average separation between the particles.

Particle number and total momentum conservation relates the
final N-particle distribution $n_N ({\bf k}_1,...,{\bf k}_N)$ 
to the initial distribution 
$n_N^0 ({\bf k}_{0,1},...,{\bf k}_{0,N})$ by
\begin{eqnarray}
   n_N ({\bf k}_1,...,{\bf k}_N)
   d^3 {\bf k}_1 ... d^3 {\bf k}_N \delta(N{\bf K}-\sum_{i=1}^N {\bf k}_i) 
   &=& n_N^0 ({\bf k}_{0,1},...,{\bf k}_{0,N})
   d^3 {\bf k}_{0,1} ... d^3 {\bf k}_{0,3} 
    \delta(N{\bf K}-\sum_{i=1}^N {\bf k}_{0,i}) \;. \nonumber\\
    &&  \label{nrel}
\end{eqnarray}
In terms of relative momenta, $q_{ij}=k_i-k_j$, and average 
four-momentum, $K=(k_1+...+k_N)/N$, we find that the
invariant momentum of N particles is
\begin{eqnarray}
   Q_N^2\equiv \frac{1}{2}\sum_{i,j=1}^N q_{ij}^2 =-(NK)^2-(Nm)^2
        =  (\sum_{i=1}^N E_i)^2 \;-\; N^2(m^2+{\bf K}^2) \,,
\end{eqnarray}
and likewise for initial momenta. With energy conservation in the Coulomb 
potential, Eq. (\ref{ec}), we obtain
\begin{eqnarray}
   Q_N^2-Q_{0,N}^2=N^2 (N-1) E_K\frac{e^2}{R_0} \,, \label{Qrel}
\end{eqnarray}
where $E_K=m_\perp\cosh Y$.
Defining the corresponding N-body correlation functions analogous to
(\ref{rel3}) we obtain from Eq. (\ref{nrel}) and (\ref{Qrel})
\begin{eqnarray}
    C_N(Q_N) = \left(\frac{Q_{0,N}}{Q_N}\right)^{3N-5}
         C^0_N (Q_{0,N})\;. \label{bbrel}
\end{eqnarray}
The measured correlation function is thus the
original one, $C^0_N(\sqrt{Q_N^2-N^2(N-1)E_Ke^2/R_0})$, scaled by powers
of $Q_{0,N}/Q_N$. In the two-body case the results reduce to those in
Ref. \cite{BB}. As an numerical example, we consider the 
three-body correlations of NA44 in LCMS (where $Y=0$) 
for pions with $\langle m_{\perp} \rangle\sim 300 $ MeV.
Assuming $R_0\sim r_0\simeq 9$ fm the three-body invariant momentum is
$Q_3^2-Q_{0,3}^2\simeq$ (30 MeV)$^2$. In comparison the lowest measured 
data point in~\cite{janus} is $Q_3 \sim 45$ MeV.

The data of NA44~\cite{rama,sanjeev} was corrected by
the Gamow factor 
\begin{eqnarray}
   C_2(q_{12}) &=& \Gamma(q_{12}) C_2^0 (q_{12}) \,,\\
   \Gamma (q_{12})&=&\frac{2\pi\eta}{\exp(2\pi\eta)-1}\,,\quad 
   \eta=e^2m/q_{12}\;.
\end{eqnarray}
The data for the three-pion correlation function~\cite{janus} was
corrected ad hoc by a three-body Gamow factor $\Gamma (q_{12})\Gamma
(q_{13})\Gamma (q_{23})$. Hereby Coulomb effects are overcorrected for
but so are they in the two-body case at the smallest measured relative
momenta . A better analysis including Coulomb effects correctly in
two- and three-body correlation functions should be performed with a
realistic charge distribution and not just a one parameter model as in
\cite{BB} and Eq. (\ref{bbrel}).

\section{Conclusion}

In this paper we develop a simple procedure to predict many--particle
correlation functions from Bose-Einstein interference between
identical particles emitted independently from an incoherent
source. Using the best available fits for the two--particle
correlation function, we calculate the higher order correlation
functions. Any deviation from this result should signal 
interesting new physics.

We apply our prescription to pion correlations. The experiment NA44
supplies a single data sample, which allows one to study two-- and
three--pion correlations, as well as two--pion--two--kaon correlations. We
test our approach against the data for two-- and three--pion correlations in
terms of the total invariant momentum of the pions and find good
agreement. In the case of three-pion correlations no additional
three--particle correlations seem to be necessary to predict the available
data. We show, how to evaluate the two-pion-two--kaon correlation function
from the currently available data. 

Larger data samples, like the one from the current Pb+Pb run at CERN
will provide us with larger data samples and allow us a more thorough 
investigation of this issue. This data sample will not only improve
the resolution for the three--pion correlation function, but also allow us
to study 4--pion correlations.

Finally, Coulomb effects were discussed and a simple N-particle generalization
of the two-particle Baym \& Braun-Munzinger correction was given in terms of
invariant momenta. 


\section*{Acknowledgements}

We would like to thank Hans B{\o}ggild, Rama Jayanti, Bengt L\"{o}rstad,
Sanjeev U. Pandey and
 Janus Schmidt--S{\o}rensen for helpfull discussions and for providing
and explaining the necessary data. 
\\

\end{document}